\documentclass{aa}  

\usepackage{graphicx}
\usepackage{txfonts}

\usepackage{lscape}
\usepackage{hyperref}
\usepackage{xcolor}
\usepackage{gensymb}
\usepackage{natbib}

\hypersetup{
	pdfnewwindow=true,  
	colorlinks=true,    
	linkcolor=blue,  
	citecolor=blue,    
	filecolor=cyan,     
	urlcolor=black,      
} 

\begin{document}

\title{A  galaxy group candidate at $z\approx3.7$ in the COSMOS field}

\titlerunning{A galaxy group at $z\approx3.7$}
\authorrunning{Sillassen et al.}

\author{
Nikolaj B. Sillassen\inst{1,2},
Shuowen Jin\inst{1,2,\thanks{Marie Curie Fellow}},
Georgios E. Magdis\inst{1,2,3}, 
Emanuele Daddi\inst{4},
John R. Weaver\inst{1,3},
Raphael Gobat\inst{5},
Vasily Kokorev\inst{1,3},
Francesco Valentino\inst{1,3},
Alexis Finoguenov\inst{6},
Marko Shuntov\inst{7},
Carlos G\'omez-Guijarro\inst{4},
Rosemary Coogan\inst{4},
Thomas R. Greve\inst{1,2},
Sune Toft\inst{1,3},
and David Blanquez Sese\inst{1,2}
          }

   \institute{Cosmic Dawn Center (DAWN)\\
      \email{sillassennikolajb@gmail.com; shuji@space.dtu.dk}
    \and
            DTU-Space, Technical University of Denmark, Elektrovej 327, DK-2800 Kgs. Lyngby, Denmark
    \and
            Niels Bohr Institute, University of Copenhagen, Jagtvej 128, DK-2200 Copenhagen, Denmark
    \and
            Universit\'e Paris-Saclay, Universit\'e Paris Cit\'e, CEA, CNRS, AIM, 91191, Gif-sur-Yvette, France
    \and
            Instituto de Física, Pontificia Universidad Católica de Valparaíso, Casilla 4059, Valparaíso, Chile
    \and
            Department of Physics, University of Helsinki, Gustaf Hällströmin katu 2, 00014 Helsinki, Finland
    \and
            Institut d'Astrophysique de Paris, UMR 7095, CNRS, and Sorbonne Universit\'e, 98 bis boulevard Arago, 75014 Paris, France
             }

   \date{Received 2 August 2022 / Accepted 2 September 2022}

 \abstract
{
We report a galaxy group candidate HPC1001 at $z\approx3.7$ in the COSMOS field. This structure was selected as a high galaxy overdensity at $z>3$ in the COSMOS2020 catalog. It contains ten candidate members, of which eight are assembled in a $10''\times10''$ area with the highest sky density among known protoclusters and groups at $z>3$. Four out of ten sources were also detected at 1.2~mm with Atacama Large Millimeter Array continuum observations. Photometric redshifts, measured by four independent methods, fall within a narrow range of $3.5<z<3.9$ and with a weighted average of $z=3.65\pm0.07$.
The integrated far-IR-to-radio spectral energy distribution yields a total UV and IR star formation rate ${\rm SFR}\approx 900~M_{\odot}$~yr$^{-1}$. We also estimated a halo mass of $\sim10^{13}~M_\odot$ for the structure, which at this redshift is consistent with potential cold gas inflow. Remarkably, the most massive member has a specific star formation rate and dust to stellar mass ratio of $M_{\rm dust}/M_{*}$ that are both significantly lower than that of star-forming galaxies at this redshift, suggesting that HPC1001 could be a $z\approx3.7$ galaxy group in maturing phase. If confirmed, this would be the earliest structure in maturing phase to date, and an ideal laboratory to study the formation of the earliest quiescent galaxies as well as cold gas accretion in dense environments.
}

\keywords{Galaxy: evolution -- galaxies: high-redshift -- submillimeter: galaxies -- galaxies: clusters}

\maketitle
 

\section{Introduction}

In the last two decades, massive galaxy protoclusters (see \citealt{Overzier2016} for a review) and groups have been discovered out to $z\sim5$ (e.g., \citealt{Pentericci2000,Daddi2009GN20,Capak2011Nature,Walter2012,Dannerbauer2014LABOCA,Daddi2017,Cucciati2018,Oteo2018cluster,Miller2018cluster_z4}). Theoretical work predicts that they play an important role in the early universe, significantly contributing to the cosmic star formation rate density (20--50\%) and forming most of their present-day stellar mass at $z>1.5$ \citep{Chiang2017cluster}.
In a simplified picture, galaxy protoclusters and groups are speculated to experience a "growing phase" at $z>3$ \citep{Shimakawa2018SW}, forming their stellar masses via rigorous star formation and growing the total gas reservoirs by accreting cold gas streams in hot media \citep{Dekel2013,Daddi2021Lya,Daddi2022Lya}. Later on, the core region of the cluster collapses, active galactic nucleus (AGN) activity takes place in cluster galaxies, and star formation declines --- the so-called maturing phase \citep{Shimakawa2018SW}. Eventually, the structure is totally collapsed, and it evolves into a virialized cluster or group of galaxies in the local Universe, is depleted of the cold gas, and has negligible star formation (e.g., \citealt{Ata2022cluster}). 

However, the evolutionary path of these massive structures is still under debate. One of the open questions is when and where the earliest maturing phase is taking place.
The virialization of the dark matter halo and the quiescence of cluster members are two critical indicators of the maturing phase. Finding the earliest collapsed dark matter halos and the first generation of quenching cluster members is key to answering this question.

In recent years, compact structures have been found with collapsed dark matter halos and cluster members that form the red sequence of passive galaxies at $z=2-2.5$ (e.g., \citealt{Wang_T2016cluster,Willis2020cluster}), indicating that their maturing phase took place at $z>2.5$.
SPT2349-56 at $z=4.3$ is so far the most distant, approximately virialized structure with a dark matter halo mass $M_{\rm halo}\sim 10^{13} M_\odot$ \citep{Miller2018cluster_z4,Hill2020cluster} and a promising progenitor of a massive cluster with $M_{\rm halo}\sim10^{15}M_\odot$ at $z=0$. However, SPT2349-56 is super-starbursting  (star formation rate $\rm SFR\gtrsim 6000~M_{\odot}~yr^{-1}$) with rich gas reservoirs ($M_{\rm gas}\geq 6\times10^{11}~M_\odot$) and it is unlikely that it harbors quenched galaxies. On the other hand, quiescent members in other dense environments have already been identified at $z\sim3$ \citep{Kubo2021quiescent,Kalita2021quiescent}, indicating that some structures start to decrease their star formation activity and enter into the maturing phase between $z=3$ and 4.

In this Letter we report a galaxy group candidate HPC1001 at $z\approx3.7$ in maturing phase. We adopt cosmology $H_0=73$, $\Omega_M=0.27$, and  $\Lambda_0=0.73$ as well as a Chabrier initial mass function \citep{Chabrier2003}.

\begin{figure*}[ht]
\setlength{\abovecaptionskip}{-0.1cm}
\setlength{\belowcaptionskip}{-0.2cm}
\centering
\includegraphics[width=0.95\textwidth]{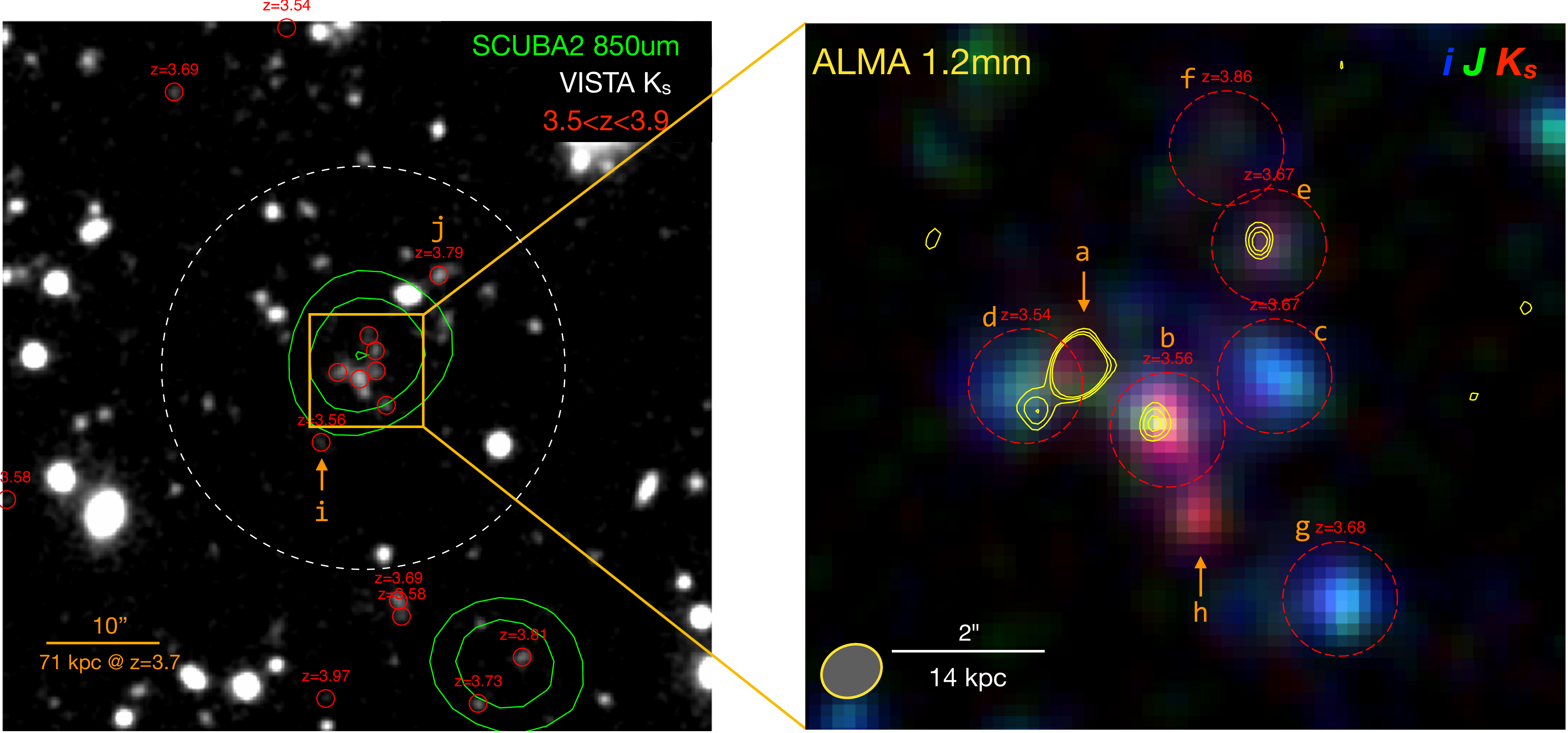}
\caption{%
The galaxy overdensity HPC1001 at $z\approx3.7$. {\it Left:} VISTA $K_s$ image overlaid with SCUBA2 850$\mu$m contours in $3$ to 5$\sigma$ levels. Galaxies at $3.5<z<3.9$ are marked with red circles and labeled with photo-z. The dashed circle shows the virial radius of a dark matter halo $M_{\rm DM}=10^{13}~M_\odot$.
{\it Right:} HSC $i$ and VISTA $JK_s$ color image for the core region, overlaid by ALMA 1.2~mm continuum contours in yellow (3 to $5\sigma$). The ALMA beam ($0.53''\times0.44''$, PA=-25.8$\degree$) is shown with a yellow ellipse in the bottom left. Candidate members are labeled with text in orange.
\label{img}
}
\end{figure*}

 \begin{figure*}[ht]
\setlength{\abovecaptionskip}{-0.1cm}
\setlength{\belowcaptionskip}{-0.2cm}
\centering
\includegraphics[width=0.92\textwidth]{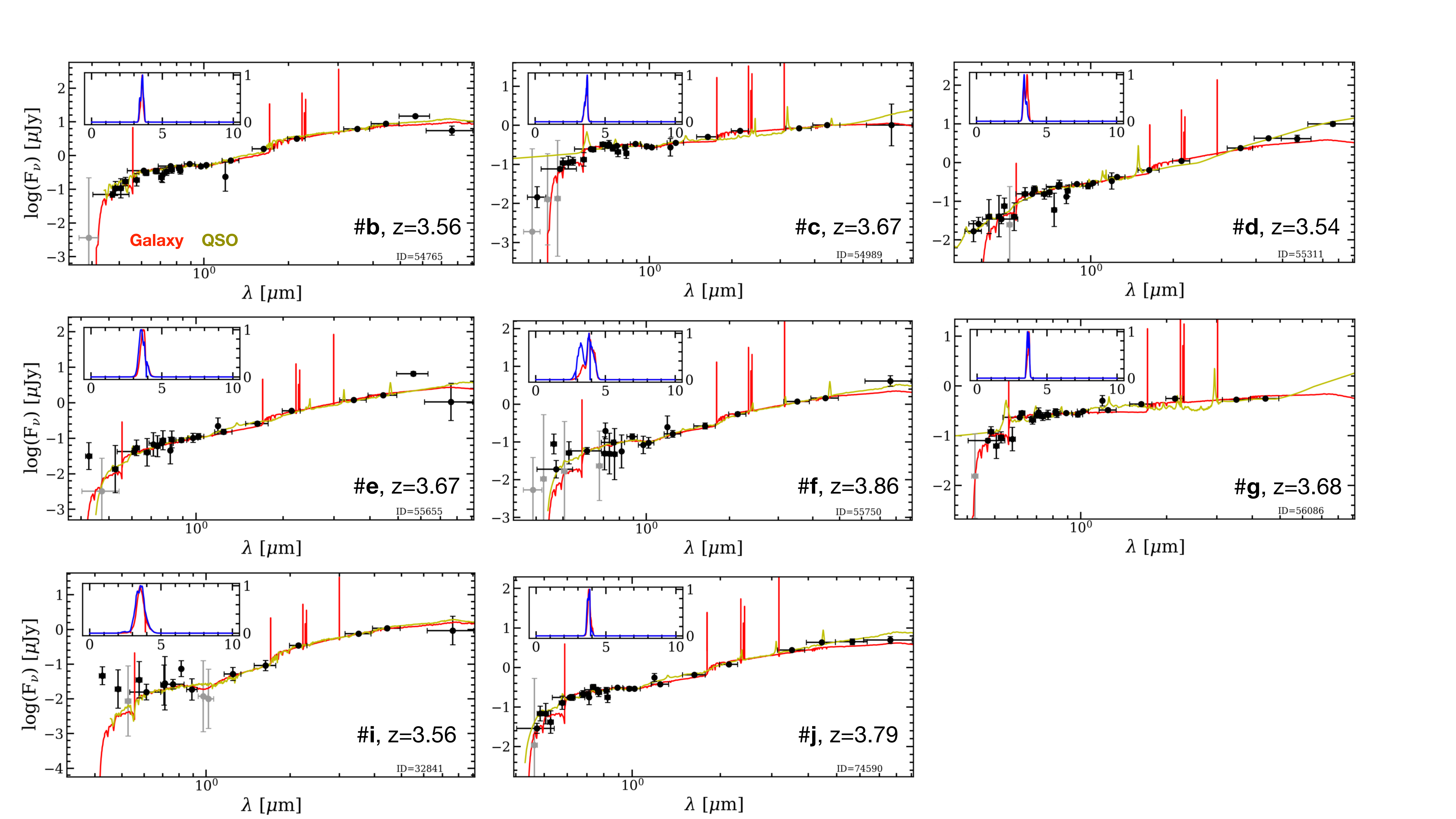}
\caption{%
Optical and NIR SEDs from the COSMOS2020 catalog. Each SED was fit by both a galaxy template (red curve) and a QSO template (yellow curve). Lines on subpanels mark the Bayesian (red) and  $\chi^2$ (blue) PDF($z$) of the galaxy template fitting, respectively. The best fit photo$-z$ estimates are also quoted in each panel.
\label{SED}
}
\end{figure*}

\section{Selection, data, and measurements}

\subsection{Selection}
Using \textsc{GalCluster}\footnote{https://github.com/Nikolaj-B-Sillassen/galcluster}, which is publicly available software designed to automatically search for galaxy overdensities, we mapped the distance of the fifth neighborhood $\Sigma_{5th}$ for $i$- and $K_s$-detected sources (${\rm S/N}_{i}>5$ \& $K_s<25$ AB mag) at $z_{\rm phot}>3$  in the COSMOS2020 catalog \citep{Weaver2022COSMOS2020}. HPC1001, which is centered at RA 150.4656$\degree$ and Dec 2.6359$\degree$, is the highest overdensity with $\Sigma_{\rm 5th}$ that is $6.8\sigma$ larger than the mean density of the COSMOS2020 catalog at $z_{\rm phot}>3$.
As shown in Fig.~\ref{img}, HPC1001 hosts a compact {group} of galaxies with $z_{\rm phot}\sim3.7$. Remarkably, the central region contains eight galaxies in a $10''\times10''$ area ($70\times70~$\,kpc$^2$, Fig.~\ref{img}-right). Comparing it to protoclusters in literature, HPC1001 has the highest sky density of galaxies at $z>3$ (Fig.~\ref{fig:a1}). 
On a large scale, HPC1001 is surrounded by near-IR- (NIR-) detected galaxies (Fig.~\ref{img}-left) at a similar redshift $3.3<z<3.9$ (475\,cMpc), and it is hosted by an overdensity of submillimeter galaxies (Fig.~\ref{fig:a2}). The density of submillimeter galaxies around HPC1001 is $\sim0.5$ arcmin$^{-2}$ for $S_{\rm 850\mu m}>5$~mJy sources, which is comparable with that in the Spiderweb protocluster at $z=2.16$ \citep{Dannerbauer2014LABOCA}.

\subsection{Optical+NIR SEDs and photometric redshifts}
We identified ten sources as candidate members within the virial radius (see Section 3.1 for the calculation), of which eight are detected in the COSMOS2020 catalog \citep{Weaver2022COSMOS2020} (Fig.~\ref{img}).
COSMOS2020 uses two different methods for obtaining photometry: the aperture photometry (Classic) and \textsc{The Farmer} and two algorithms, \textsc{LePhare} and \textsc{EAZY} \citep{Ilbert2006LePhare,Brammer2008EAZY}, for the spectral energy distribution (SED) fitting of each of the two versions of photometry. Thus, for the eight sources that are included in the COSMOS2020 catalog, we have four independent photo$-z$ estimates that fall within a narrow range $3.5<z<3.9$ (306\,cMpc, Fig.~\ref{fig:b1}) and for each source are consistent within uncertainties.
Six of these eight galaxies are also included in the COSMOS2015 catalog \citep{Laigle2016}, with photo$-z$ consistent with those derived in our work based on the COSMOS2020 photometry (Fig.~\ref{fig:b1}). 
We note though that the COSMOS2020 Classic catalog (and the COSMOS2015 catalog) is using aperture photometry of $2''$ ($3''$), which is severely affected by blending in crowded regions such as HPC1001. On the other hand, Farmer photometry is extracted by simultaneous modeling of the profiles of all galaxies within the group, making it more robust against blending. Furthermore, we calculated the systematic redshift bias at $3.5<z<4$ by cross-matching the COSMOS spec-$z$ catalog (M. Salavato et al. in prep.) to the Farmer \textsc{LePhare} photo-$z$, and found it to be negligible, that is to say median $(z_{\rm phot}-z_{\rm spec})/(1+z_{\rm spec})=1.3\times10^{-3}$. 
Therefore, we chose to adopt the COSMOS2020/Farmer - \textsc{LePhare} combination as the benchmark for our analysis, in which the average photo$-z$ uncertainty of this sample is 0.2.

In Fig.~\ref{SED} we present the optical and NIR SEDs for the eight sources, and list the derived photo-$z$, stellar mass estimates in Table~\ref{tab:1}.
Apart from a galaxy template, we also performed a fit assuming a quasar template to examine the systematics of the galaxy template solutions (but not to serve as a bona fide AGN classification).
A galaxy template provides the best fit for seven out of eight sources, ensuring the fidelity of the derived stellar masses. 
The SEDs of HPC1001.d and f show a blue excess to the galaxy template solution, suggesting potential AGN activity.
As a sanity check, we also compared the redshift probability distribution functions (PDFs) as derived based on Bayesian and $\chi^2$ methods and were found to be consistent (Fig.~\ref{SED}). 

The remaining two sources, HPC1001.a and h, are not included in COSMOS2020 catalog, as they are only detected in $K_s$ while their signal-to-noise ratios (S/Ns) are lower than the detection criteria in the $JHK_s$-stacked image adopted by the COSMOS2020 analysis. However, HPC1001.h is included in the COSMOS2015 catalog with $z_{\rm phot}=3.79\pm0.4$ (Table~\ref{tab:1}).

In order to properly deblend HPC1001, we repeated the \textsc{Farmer} photometry this time by adding HPC1001.a and h in the prior catalog (at their respective ALMA position for a and $K_s$ position for h), and by simultaneously fitting all sources in the $K_s$ and IRAC images. The new photometry was used to derive the properties of HPC1001.a while it remained virtually unchanged for the eight sources that were included in the original COSMOS2020 catalog. The SED of HPC1001.h is highly uncertain due to its nondetection at wavelengths shorter than $K_s$ and its faintness in IRAC ($M_{\rm AB}>28$). While in Table~\ref{tab:1}, we list its parameters as derived in the COSMOS2015 catalog, we caution that its photo-$z$ and $M_*$ are only indicative and their uncertainties could be underestimated.

\begin{figure*}[ht]
\setlength{\abovecaptionskip}{-0.1cm}
\setlength{\belowcaptionskip}{-0.2cm}
\centering
\includegraphics[width=0.48\textwidth]{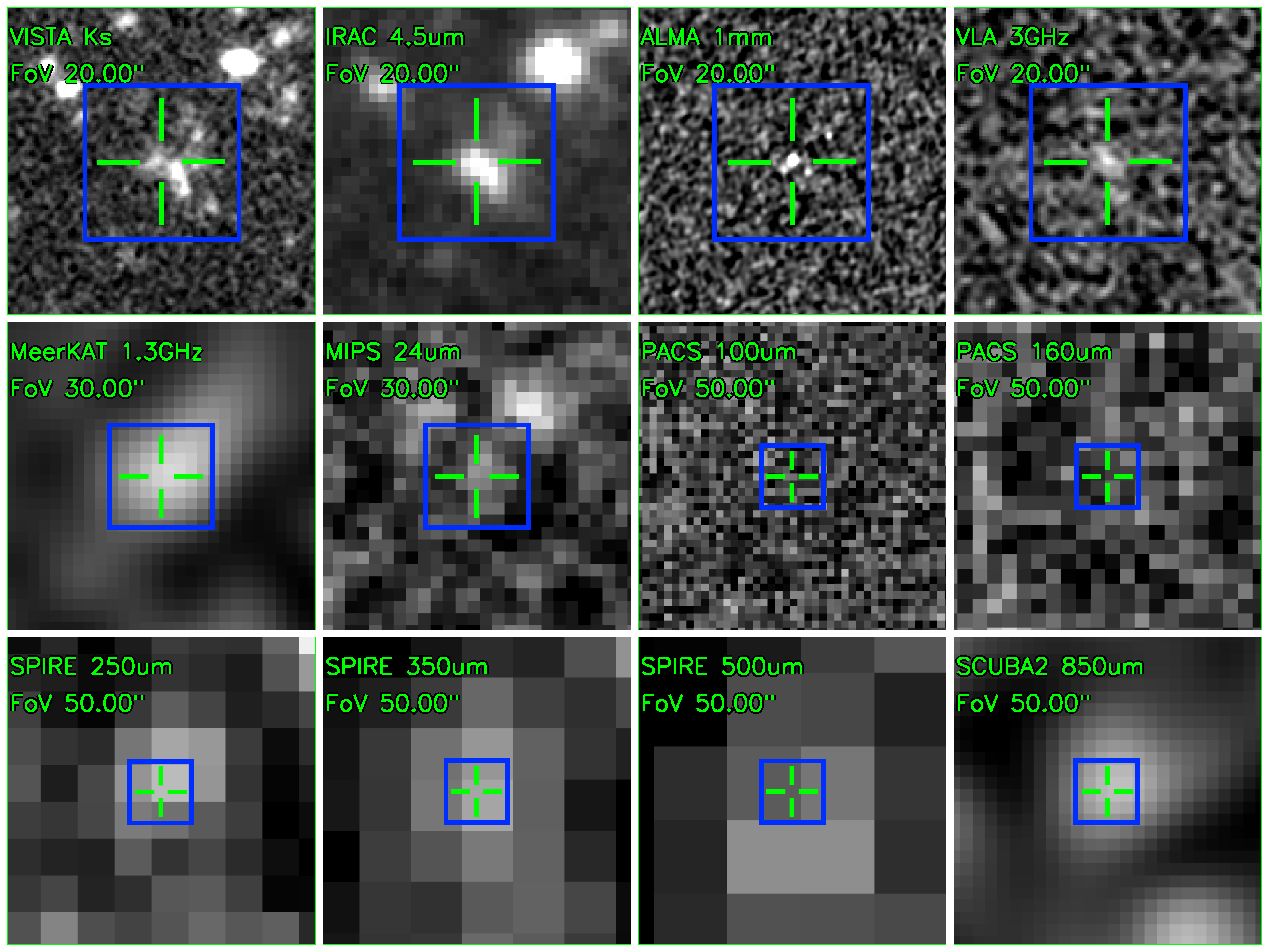}
\includegraphics[width=0.48\textwidth,trim={0.0cm 0.0cm 0.0cm 0.0cm}, clip]{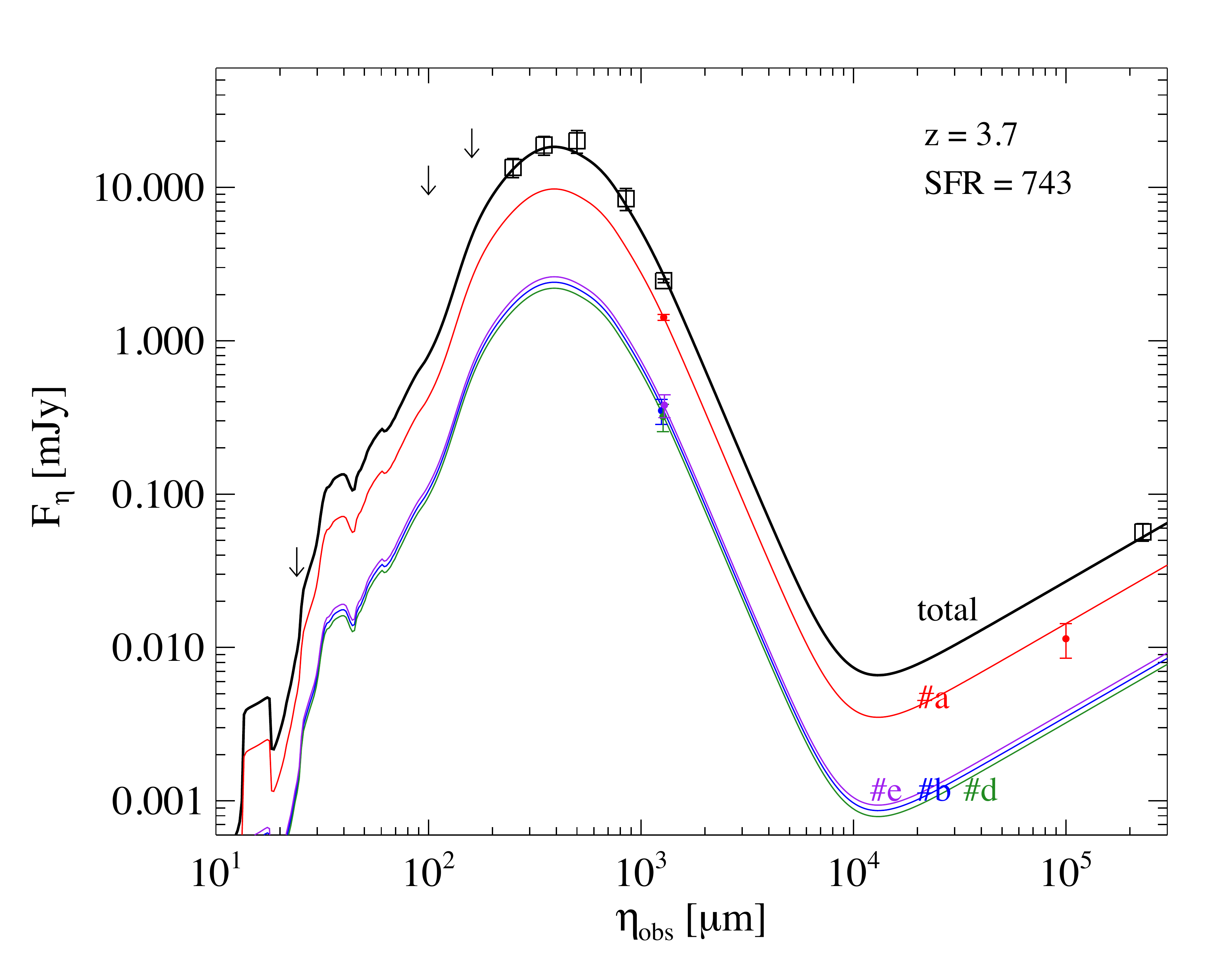}
\caption{%
FIR-to-radio properties of HPC1001.
{\it Left:} Multiband images of HPC1001. The instrument, wavelength, and field of view (FoV) are marked in green text. The blue box marks a 10"$\times$10" square.
{\it Right:} Integrated and individual FIR, millimeter, and radio SEDs, fitted at $z=3.7$ by a starbursting template \citep{Magdis2012SED} and a FIR-related radio component \citep{Magnelli2013}.  
\label{cutout}
}
\end{figure*}

\subsection{FIR, (sub)millimeter, and radio SED}
HPC1001 is covered by Atacama Large Millimeter Array (ALMA) Band6 continuum imaging (Fig.~\ref{img}-right), as part of the project ID: 2013.1.00034.S (PI: N. Scoville). While no lines were identified in the data cubes, continuum emission at 1.2~mm was detected from four sources at 5--23$\sigma$ significance levels. Among the four ALMA sources, only HPC1001.a is slightly resolved, while the rest are point-like. We thus adopted aperture photometry for HPC1001.a and peak fluxes for HPC1001.b, d, and e.

HPC1001 was also detected (but barely resolved) in Herschel, SCUBA2 \citep{Simpson2019}, and MeerKAT \citep{Jarvis2016mightee} imaging (Fig.~\ref{cutout}-left). We measured its integrated far-IR (FIR) to radio photometry by performing the super-deblending technique as described in \cite{Jin2018cosmos} at the fixed position of HPC1001.a.
The Herschel colors ($S_{\rm 250\mu m}<S_{\rm 350\mu m}<S_{\rm 500\mu m}$) indicate that it is a 500$\mu$m riser which is even redder than the cluster CL J1001 at $z=2.5$, supporting its $z>2$ nature   \citep{Riechers2017,Donevski2018red,Cairns2022}.

In order to constrain the IR star formation rate (SFR) and dust mass ($M_{\rm dust}$) of the group, we fit the integrated FIR to radio photometry using the \textsc{StarDust} code \citep{Kokorev2021}.
The integrated SED (Fig.~\ref{cutout}-right) was well fitted by a starburst template (GN20, \citealt{Magdis2012SED}) at $z=3.7$ with dust-obscured ${\rm SFR_{IR}}=743~M_\odot~yr^{-1}$. 
Assuming that all ALMA detected sources share the same  SED shape as the integrated SED, we derived SFRs and dust masses for the four ALMA sources by splitting the total values based on the ALMA 1.2~mm flux of each source (Fig.~\ref{cutout}-right and Table~\ref{tab:1}). 
Given the ALMA brightness and NIR faintness, HPC1001.a could be a starburst with optically thick dust emission, in which case the inferred dust mass is likely to be overestimated by a factor of two \citep{Cortzen2020GN20,Jin2022}. This consideration has been incorporated in the adopted dust mass uncertainty of the source (Table~\ref{tab:1}).

\begin{table*}[ht]
{
\caption{Physical properties of HPC1001}
\label{tab:1}
\renewcommand\arraystretch{1.25}
\centering
\begin{tabular}{ccccccccccc}
\hline\hline
     Name       &  ID$^a$   &  $z_{\rm phot,NIR}$ & $S_{\rm 1.2mm}$ & $\log(M_{*}/M_{\odot})$ & ${\rm SFR_{UV,cor}}$  &  ${\rm SFR_{IR}}$ & $\log(M_{\rm dust}/M_\odot)$  & $f_{\rm gas}$\\
                 &         &   &   [mJy]  &  &    [M$_\odot/$yr]        &  [M$_\odot/$yr] &   &  [$M_{\rm gas}/M_{*}$] \\
 \hline
HPC1001.a  & --    &   3.7$^b$  &  1.42$\pm$0.06  & 10.3$^{+0.2}_{-0.2}$   & --  &  429$\pm$20 &  8.93$^{+0.05}_{-0.35}$ & 4.3$^{+2}_{-4}$ \\
HPC1001.b & 54765 &    3.56$^{+0.08}_{-0.10}$  & 0.35$\pm$0.06   &   11.0$^{+0.1}_{-0.1}$     & 110$^{+25}_{-66}$    & 106$\pm$20 & 8.32$\pm$0.08 & 0.2$\pm$0.1 \\
HPC1001.c &  54989  &  3.67$^{+0.07}_{-0.14}$ & $<$0.18   &  9.8$^{+0.1}_{-0.1}$   &   17$^{+3}_{-8}$   & <59 & <8.07 & <1.9\\
HPC1001.d & 55311  &   3.54$^{+0.13}_{-0.17}$ &  0.32$\pm$0.06  &   10.5$^{+0.1}_{-0.1}$   & 64$^{+12}_{-11}$   & 97$\pm$20 & 8.28$\pm$0.09  & 0.6$\pm$0.2\\
HPC1001.e & 55655   &   3.67$^{+0.19}_{-0.18}$  &  0.38$\pm$0.06   &   10.2$^{+0.1}_{-0.1}$  & 50$^{+46}_{-35}$   & 115$\pm$20 & 8.36$\pm$0.09 & 1.4$\pm$0.5\\
HPC1001.f & 55750   &    3.86$^{+0.32}_{-0.39}$  & $<$0.19   & 10.2$^{+0.1}_{-0.2}$  & 32$^{+29}_{-15}$    & <61 & <8.07  & <0.7 \\
HPC1001.g & 56086 &    3.68$^{+0.06}_{-0.20}$  & $<$0.19   &   9.5$^{+0.1}_{-0.1}$     & 13$^{+2}_{-3}$  & <61 & <8.07  & <3.7 \\
HPC1001.h & 880059$^c$ &    3.79$^{+0.37}_{-0.34}$  & $<$0.18   &   10.0$^{+0.2}_{-0.2}$     & 77$^{+65}_{-42}$  & <59 & <8.07  & <0.9 \\
HPC1001.i & 32841 &    3.56$^{+0.29}_{-0.26}$  & $<$0.34   &   10.2$^{+0.1}_{-0.2}$     & 29$^{+20}_{-16}$  & <106 & <8.33  & <1.3 \\
HPC1001.j & 74590 &    3.79$^{+0.12}_{-0.10}$  & $<$0.33   &   10.6$^{+0.1}_{-0.1}$     & 73$^{+15}_{-57}$  & <126 & <8.31 &  <0.5 \\
\hline
(total)\\
HPC1001 & --  &    3.65$\pm$0.1  & 2.46$\pm$0.05   & 11.4$\pm{0.2}$  & 391$^{+66}_{-97}$ $^d$    &  743$\pm$3 & 9.17$\pm$0.05 & --\\
\hline
\end{tabular}\\
}
{Notes: $S_{\rm 1.2mm}$: ALMA 1.2~mm photometry. ${\rm SFR_{UV,cor}}$: UV-corrected SFR from the COSMOS catalogs. $^a$ COSMOS2020 Farmer catalog ID. $^b$ We adopted $z=3.7$ from the weighted average of the group. $^c$ {This source is not in the COSMOS2020 catalog, and we list parameters from the COSMOS2015 catalog instead.} 
$^d$ We did not account for HPC1001.h due to its inconsistency between SFR$_{\rm UV,cor}$ and SFR$_{\rm IR}$.
} 
\end{table*}

\begin{figure*}[ht]
\setlength{\abovecaptionskip}{-0.1cm}
\setlength{\belowcaptionskip}{-0.3cm}
\centering
\includegraphics[width=0.32\textwidth]{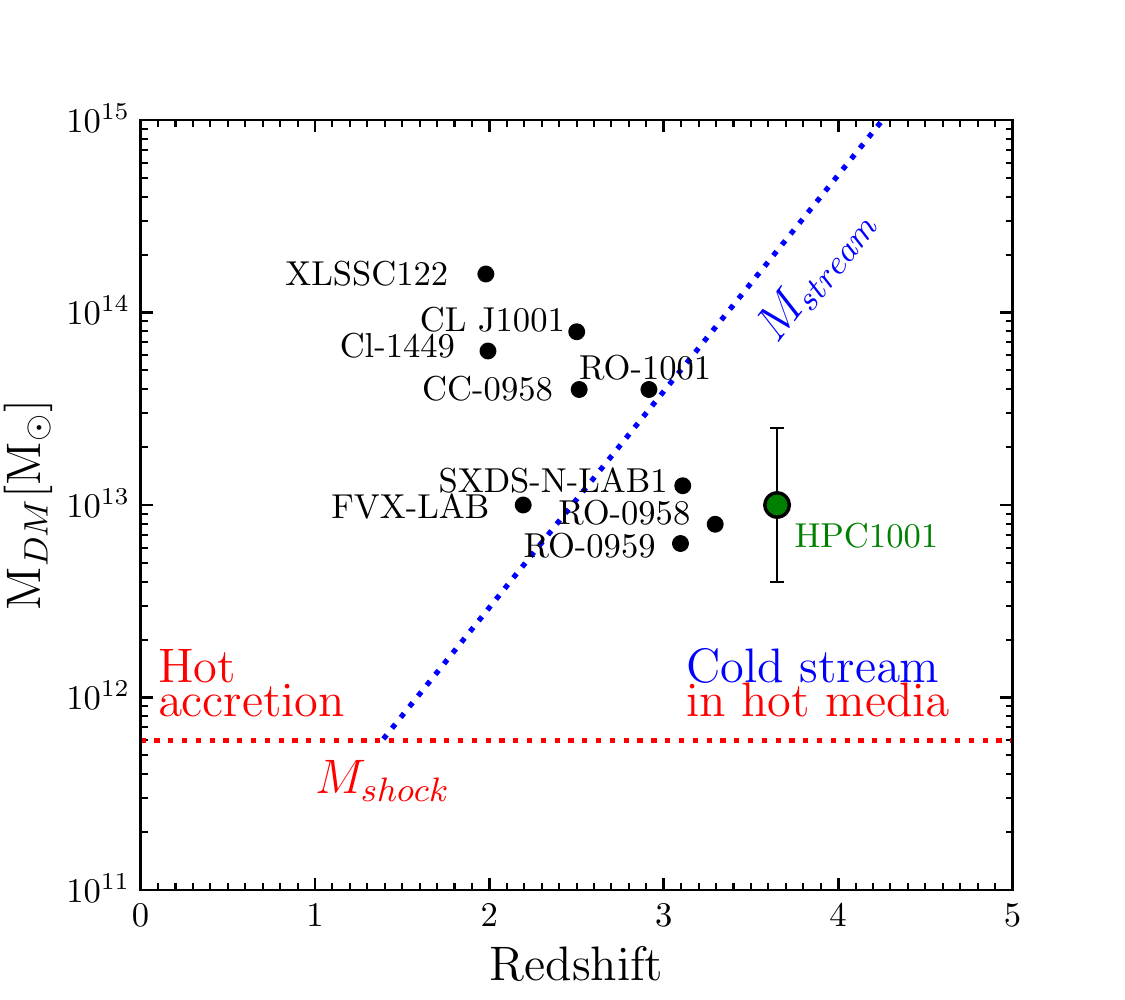}
\includegraphics[width=0.27\textwidth]{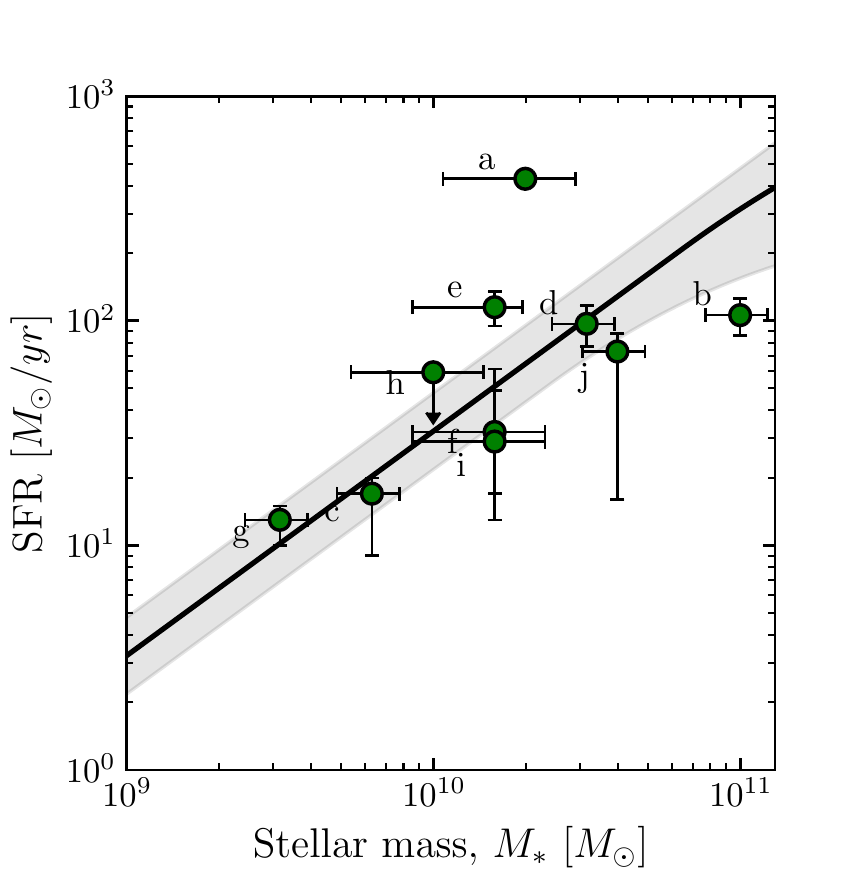}
\includegraphics[width=0.4\textwidth]{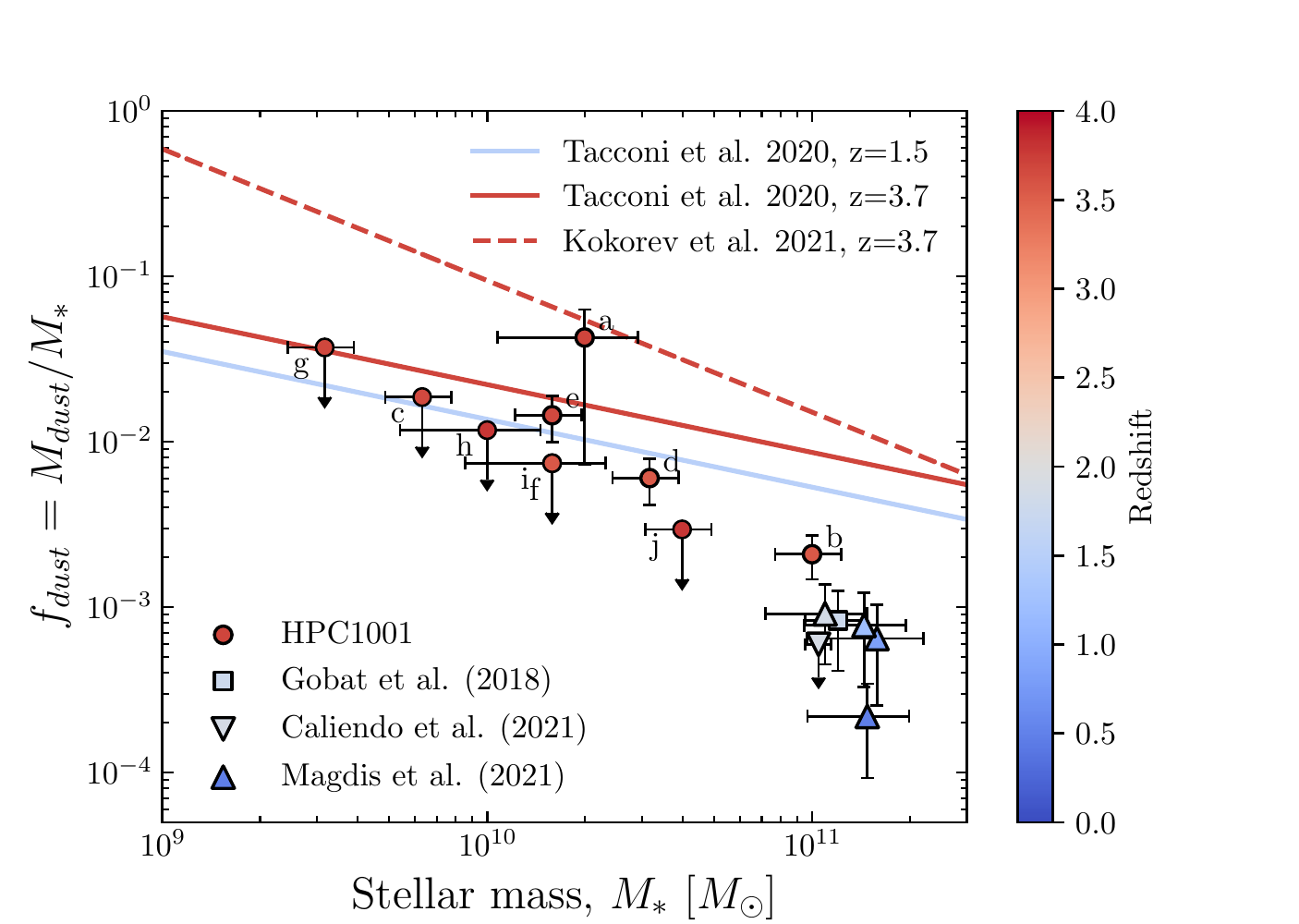}
\caption{
Physical properties of HPC1001.
{\it Left:} Dark matter halo mass for (proto)clusters and groups in literature and this work. Blue- and red-dashed lines show critical masses of the cold stream and shock from simulations in \cite{Dekel2013}.
{\it Middle:} SFR versus stellar mass. The solid line with a shaded area indicate the star-forming main sequence (MS) at $z=3.7$ and its $1\sigma$ scatter \citep{Schreiber2015}, respectively.
Individual sources in HPC1001 are shown in green filled circles with their ID labeled with text.
{\it Right:} Dust to stellar mass ratio $M_{\rm dust}/M_{\ast}$ as a function of stellar mass. Main sequence scaling relations are color-coded in redshift. Dashed and solid lines show main sequences from  \cite{Kokorev2021} and \cite{Tacconi2020}, respectively. 
The square and triangle show quiescent galaxy samples in literature.
HPC1001.b shows comparably low $f_{\rm dust}$ as quiescent galaxies at $z\sim1.5$ with respect to MS.
\label{halo}
}
\end{figure*}

\section{Results}

\subsection{Dark matter halo mass}

We estimated the dark matter halo mass $M_{\rm DM}$ of HPC1001 with six methods: (1) Using the $M_{\rm halo}$-$M_{\ast}$ scaling relation from \cite{Behroozi2013Mhalo} and the stellar mass of the most massive source HPC1001.b, it yields a lower limit for the halo mass of log$(M_{\rm DM}/M_\odot)\gtrsim13$; (2) following the methodology presented in \cite{Daddi2021Lya,Daddi2022Lya}, we estimated the halo mass based on the stellar masses above the completeness limit of the COSMOS survey, that is $\log(M_*/M_{\odot})>9.7$ at $z=3.7$ \citep{Weaver2022COSMOS2020}. This mass limit holds for nine out of ten members, and the sum of stellar masses is $M_*=2.45\times10^{11}~M_\odot$. We then extrapolated a total stellar mass of $M_{\rm *,total}=2.6\times10^{11}~M_\odot$
down to $10^7M_{\odot}$,  assuming the stellar mass function of field galaxies at $3<z<4$  \citep{Muzzin2013}. Adopting the dynamical mass-constrained  $M_{\rm halo} - M_{\ast}$ scaling relation for $z\sim1$ clusters with $0.6 \times 10^{14} <M/M_{\odot}<16\times 10^{14}$ \citep{van_der_Burg2014} yields
a halo mass of $\log(M_{200}/M_{\odot})$ =12.8; 
(3) adopting the stellar-to-halo mass relation of \cite{Shuntov2022} and $M_{\rm *,total}=2.6\times10^{11}~M_\odot$, we obtained a halo mass of log$(M_{\rm DM}/M_\odot)=12.7$;
(4) we used the background and point source subtracted combined Chandra+XMM-Newton image in the 0.5-2 keV \citep{Gozaliasl2019Xray} to study the X-ray emission of HPC1001. No extended source was detected at the position of the group. Using a $32''$ radius aperture, we placed a two sigma upper limit $S_{\rm 0.5-2 keV} < 4.2\times 10^{-16}~{\rm ergs~s^{-1}~cm^{-2}}$, which yielded an upper limit of $M_{\rm 200}<2.9\times10^{13}~M_\odot$;
(5) assuming a group velocity dispersion $\sigma_V=400$\,km~s$^{-1}$, we found that the galaxy number density of HPC1001 (in putative $R_{\rm vir}<20''$) is more overdense than the average by a factor of 180 at $z\sim3.7$ in COSMOS2020. Applying a mean baryon and dark matter density of $7.77\times10^{-26}~$kg~m$^{-3}$ in comoving volume from Planck cosmology and a galaxy bias factor of 10--20 at $z=3.7$ \citep{Tinker2010haloBias}, we obtained a halo mass of log$(M_{\rm DM}/M_\odot)=13-13.4$; 
(6) assuming that HPC1001 is the most extreme halo in COSMOS at $z>3.65$, that is to say that it is the only case that can be found in a 2 deg$^2$ field, the halo mass would be $M_{\rm DM}\sim2.7\times10^{13} M_\odot$ in Planck cosmology.

Based on these estimates, we adopted an average halo mass of log$(M_{\rm DM}/M_\odot)=13$ with an uncertainty of 0.4\,dex that is representative at these faint levels (e.g., \citealt{Looser2021,Daddi2022Lya}). Assuming the structure is collapsed, the virial radius corresponding to this halo mass would be 140 kpc \citep{Goerdt2010core}, which is large enough to enclose all ten sources that are included in deriving the halo mass.

In Fig.~\ref{halo}-left, we compared the halo mass of HPC1001 to literature data taken from \cite{Daddi2022Lya}. With a halo mass that is lower than that of massive clusters, such as CL J1001, but comparable to that of galaxy groups, such as RO-0958 at $z=3.3$ and RO-0959 at $z=3.1$ \citep{Daddi2022Lya}, we propose that HPC1001 is likely to be a galaxy group. Its halo mass is above the mass criterion for generating shocks, and it is well within the region where cold streams are accreting in hot media, as predicted by simulations \citep{Dekel2013}. In this case, cold gas inflow is taking place in HPC1001 and similar to RO-1001 and RO-0959 \citep{Daddi2021Lya,Daddi2022Lya}, this inflow should be detectable via diffuse Ly$\alpha$ emission.

\subsection{Star formation activity \& dust fraction}

In order to obtain the total SFR within the HPC1001 structure, we summed the  ${\rm SFR_{IR}}$ of the four ALMA-detected sources with the ${\rm SFR_{UV,cor}}$ of the remaining unobscured sources. The total UV+FIR star formation rate is ${\rm SFR_{total}\approx900~M_\odot~yr^{-1}}$. 
Applying the scaling relations of \cite{Daddi2022Lya}, for a dark matter halo with log$(M_{\rm DM}/M_\odot)=13$ at $z=3.7$, the expected cold baryonic accretion rate (BAR) is 2780 M$_\odot$ yr$^{-1}$ which in turn corresponds to a SFR of 800 M$_\odot$ yr$^{-1}$. The total SFR of HPC1001 agrees with the model prediction, further supporting the scenario that HPC1001 is a cold-accretion-fed group.

Moving the focus to individual sources, we find that HPC1001.b has a specific star formation rate (sSFR) that places the source $\times 3.0\pm1.0$ below the main sequence at the corresponding redshift (Fig.~\ref{halo}-middle). At the same time, as shown in Fig.~\ref{halo}-right, its dust to stellar mass ratio $f_{\rm dust}$ is lower by a factor of $\times4-7$ with respect to that of main sequence galaxies at $z=3.7$ \citep{Tacconi2020,Kokorev2021} and comparable to that of $z\sim1.5$ quiescent galaxies \citep{Gobat2018,Magdis2021,Caliendo2021}.
Assuming $M_{\rm gas}=100\times M_{\rm dust}$ and accounting for a 20\% uncertainty due to unknown metallicity \citep{Jin2019alma,Jin2022}, HPC1001.b would have an inferred gas fraction of $M_{\rm gas}/M_{*}=0.2\pm0.1$ and a gas depletion time of 200\,Myr.

\section{Discussion}

\subsection{A galaxy group in maturing phase at $z\approx 3.7$}

Multiple pieces of evidence support the idea that HPC1001 is a galaxy group at $z\approx3.7$: (1) the high sky overdensity of galaxies; (2) the excellent agreement of photometric redshifts from four independent methods; and (3) the fact that the inferred dark matter halo mass and SFR are comparable with that of galaxy groups at $z>3$ \citep{Daddi2022Lya}.
We note that the large structure surrounding HPC1001 hosts extra NIR-detected galaxies at a similar redshift ($3.3<z<4$) that are coupled with an overdensity of submillimeter sources (Fig.~\ref{fig:a2}). 
These sources could be in subhalos that associate with HPC1001 in a megaparsec-scale structure \citep{Koyama2013cluster,Cucciati2018,Jin2021cluster}, which needs to be identified by future spectroscopic follow-ups.

On the other hand, the remarkably low gas and dust fractions of HPC1001.b suggest that this galaxy is more evolved compared to normal star-forming galaxies at the same epoch \citep{Gomez-Guijarro2019}.
These properties are in line with the maturing phase of protoclusters \citep{Shimakawa2018SW}, where the core hosts both starbursting and quenching members, the overall star formation activity begins to decline, and the dark matter halo starts to collapse. This scenario has not been observed at $z>3$, and if confirmed HPC1001 would be the earliest structure in maturing phase detected to date.

\subsection{A candidate for a galaxy in quenching}

HPC1001.b is the most massive galaxy in the structure, with a stellar mass comparable to that of the quiescent cluster members at $z\sim3$ \citep{Kubo2021quiescent,Kalita2021quiescent}. Assuming no gas inflow, HPC1001.b would deplete its molecular gas reservoir in 200~Myr and become quiescent at $z>3.2$ with $M_{*}\sim1.2\times10^{11} M_{\odot}$, forming the first generation of massive quiescent galaxies in dense environments. 
Of course, potential cold gas inflow (Fig.~\ref{halo}-left) can replenish the star-forming gas reservoir and maintain star formation in HPC1001.b. However, the presence of a quenched galaxy in the RO-1001 group where gas inflow is present \citep{Kalita2021quiescent} suggests that quiescent galaxies can form in high$-z$ dense environments despite the ongoing gas inflow.

In order to fully characterize the nature of HPC1001 and explore the quiescence of HPC1001.b, [CII]158$\mu$m and [CI]609,369$\mu$m observations are  crucial to infer spectroscopic redshifts, confirm group members, and robustly measure gas masses and SFRs. The kinematic information from [CII] or [CI] can also verify if HPC1001 is compatible with a single virialized structure. On the other hand, VLT/MUSE and GTC/MEGARA can be employed to detect the cold gas inflow via Ly$\alpha$ observations. Unfortunately, HPC1001 is not covered by the JWST COSMOS-Web survey. However, a future follow-up observation of the structure with JWST/NIRCam and NIRSpec will be a powerful tool to reveal potential inter-cluster light and measure the  metallicity of the group members.

\begin{acknowledgements}
This paper makes use of the ALMA data: ADS/JAO.ALMA 2013.1.00034.S. 
ALMA is a partnership of ESO (representing its member states), NSF (USA), and NINS (Japan), together with NRC (Canada), MOST and ASIAA (Taiwan), and KASI (Republic of Korea), in cooperation with the Republic of Chile. The Joint ALMA Observatory is operated by ESO, AUI/NRAO, and NAOJ.
The Cosmic Dawn Center (DAWN) is funded by the Danish National Research Foundation under grant No. 140.
SJ is supported by the European Union's Horizon research and innovation program under the Marie Sk\l{}odowska-Curie grant agreement No. 101060888.
SJ and GEM acknowledge financial support from the Villum Young Investigator grant 37440 and 13160.
TRG acknowledges support from the Carlsberg Foundation (Grant no CF20-0534).
\end{acknowledgements}

\bibliographystyle{aa}
\bibliography{biblio}


\begin{appendix}
\section{ Galaxy overdensity}

\begin{figure}[h]
\setlength{\abovecaptionskip}{-0.1cm}
\setlength{\belowcaptionskip}{-0.3cm}
\centering
\includegraphics[width=0.42\textwidth]{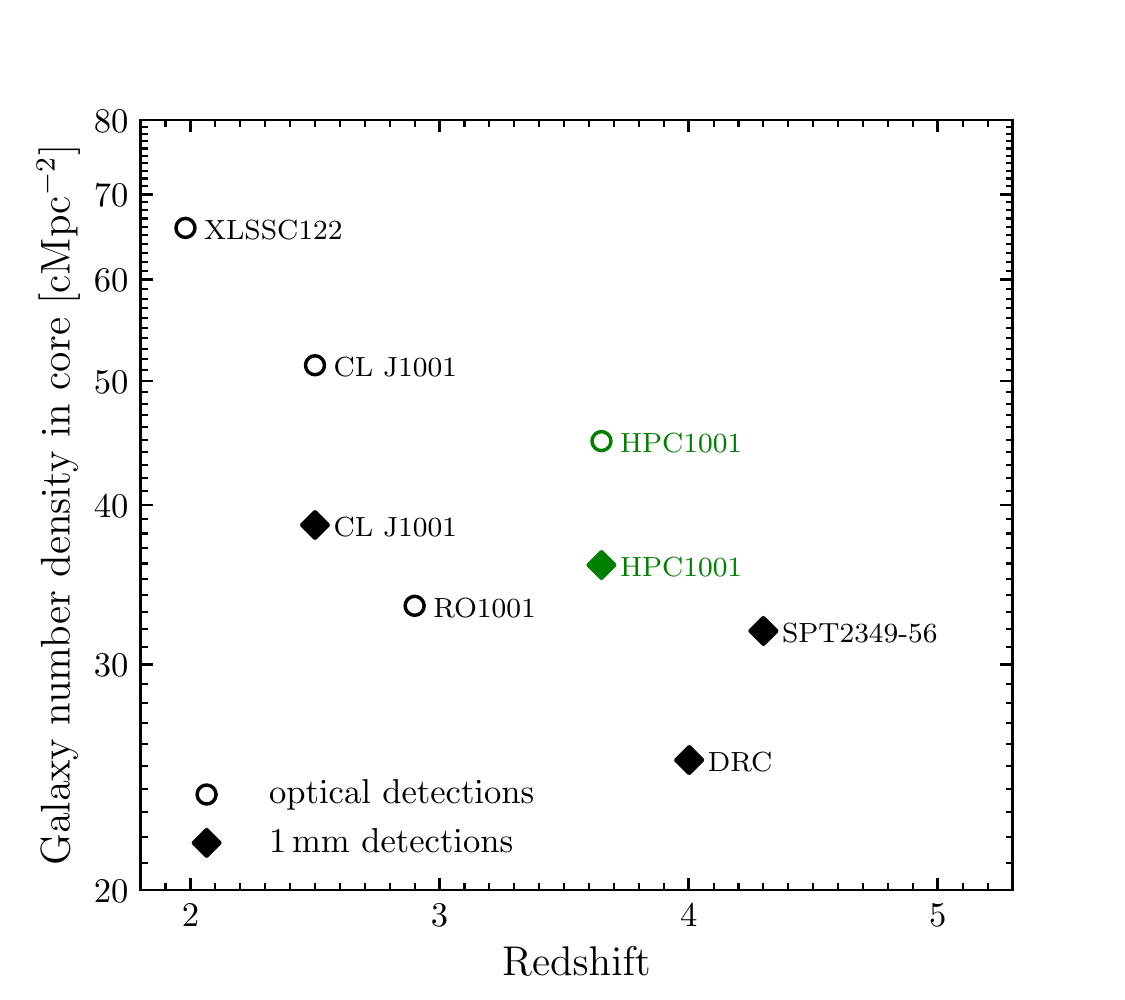}
\caption{
Galaxy sky density of (proto)cluster cores in comoving area. Densities were measured in a $10''\times10''$ area with the maximum number of cluster members covered by the area, where we applied a stellar mass limit log$(M_*/M_\odot)>10.2$ for optical-detected members and a flux limit $S_{\rm 1mm}>0.3$~mJy for 1\,mm continuum detections, respectively.
We show extreme cases in literature from \citet{Wang_T2016cluster}, \citet{Miller2018cluster_z4}, \citet{Willis2020cluster}, and \citet{Daddi2021Lya}. We note that CL~J1001 and DRC lack 1\,mm observations; their 1\,mm fluxes were derived by scaling their 870\,$\mu$m and 2\,mm continuum under assumption of a GN20 dust template \citep{Magdis2012SED}.
HPC1001 is highlighted in green, showing the highest density at $z>3$.}
\label{fig:a1}
\end{figure}

\begin{figure}[h]
\setlength{\abovecaptionskip}{-0.1cm}
\setlength{\belowcaptionskip}{-0.3cm}
\centering
\includegraphics[width=0.42\textwidth]{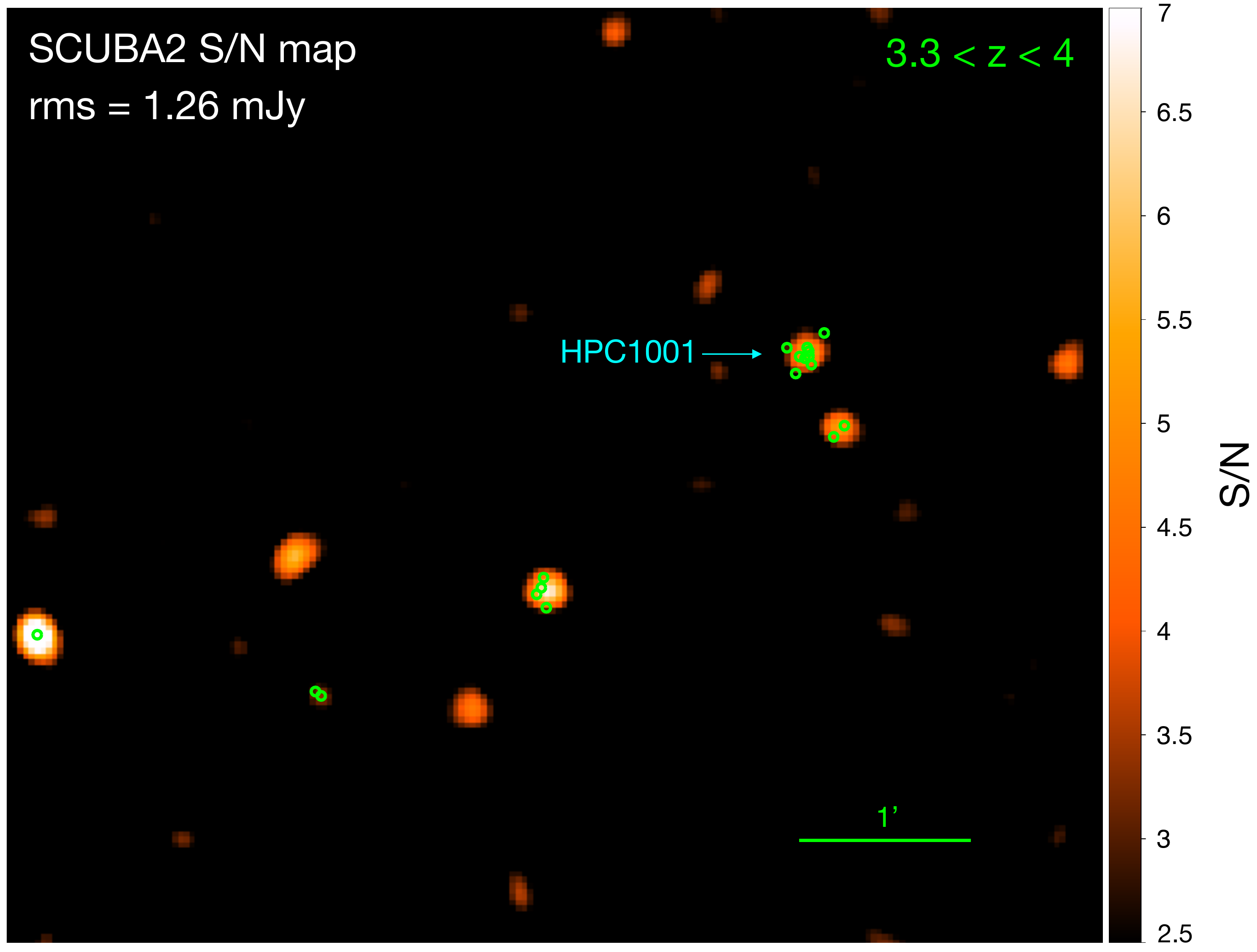}
\caption{
SCUBA2 850$\mu$m S/N map. The rms is shown with text and the S/N is indicated by the color bar from 2.5 to 7$\sigma$. Green circles mark galaxies that have photo-z $3.3<z<4$ and match SCUBA2 detections (S/N$_{850}>$3 and tolerance$<10''$). HPC1001 is in an overdensity of submillimeter galaxies at a similar redshift. 
\label{fig:a2}
}
\end{figure}

\vspace{3cm}
\section{Redshift comparison}

\begin{figure}[h]
\setlength{\abovecaptionskip}{-0.1cm}
\setlength{\belowcaptionskip}{-0.2cm}
\centering
    \includegraphics[width=0.95\linewidth]{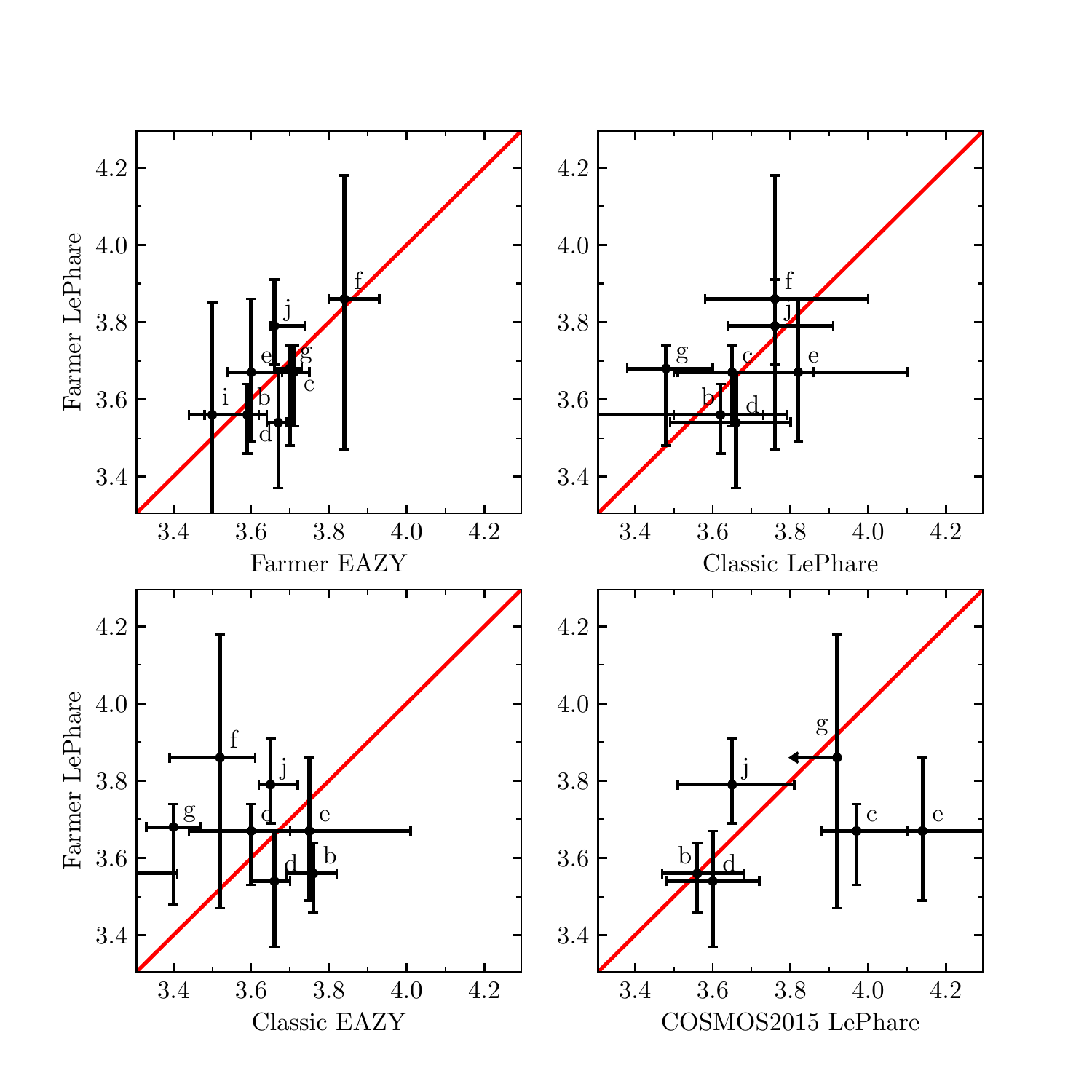}
    \caption{Comparison of photo-z between the four versions in the COSMOS2020 catalog \citep{Weaver2022COSMOS2020} and the COSMOS2015 catalog \citep{Laigle2016}. 
The weighted averages are $z=3.65\pm0.07$ (\textsc{Farmer LePhare}), $z=3.68\pm0.02$ (\textsc{Farmer} \textsc{EAZY}), $z=3.63\pm0.08$ (\textsc{Classic LePhare}), $z=3.62\pm0.04$ (Classic EAZY), and $z=3.72\pm0.08$ (COSMOS2015), respectively.}
    \label{fig:b1}
\end{figure}
\end{appendix}

\end{document}